\documentclass{article}

\usepackage{spconf,amsmath,graphicx}
\usepackage{subcaption}
\usepackage{multirow}
\usepackage{multicol}
\usepackage{graphicx}
\usepackage{microtype}
\usepackage{graphicx}
\usepackage{booktabs} 
\usepackage{amssymb}
\usepackage{hyperref}
\usepackage{threeparttable, multirow}
\usepackage{tabularx}

\usepackage{mathtools}

\newcommand{\bp}[1]{\left(#1\right)}
\newcommand{\R}{\mathbb{R}}

\DeclareMathOperator{\softmax}{softmax}
\DeclareMathOperator{\crossattention}{CrossAttention}

\def\L{{\cal L}}

\title{SLiCK: Exploiting Subsequences for Length-Constrained Keyword spotting}
\name{Kumari Nishu, Minsik Cho, Devang Naik}
\address{Apple}
\begin{document}
%
\maketitle

\begin{abstract}

 User-defined keyword spotting on a resource-constrained edge device is challenging. However, keywords are often bounded by a maximum keyword length, which has been largely under-leveraged in prior works. Our analysis of keyword-length distribution shows that user-defined keyword spotting can be treated as a length-constrained problem, eliminating the need for aggregation over variable text length. This leads to our proposed method for efficient keyword spotting, SLiCK (exploiting Subsequences for Length-Constrained Keyword spotting). We further introduce a subsequence-level matching scheme to learn audio-text relations at a finer granularity, thus distinguishing similar-sounding keywords more effectively through enhanced context. In SLiCK, the model is trained with a multi-task learning approach using two modules: \textbf{Matcher} (utterance-level matching task, novel subsequence-level matching task) and \textbf{Encoder} (phoneme recognition task). The proposed method improves the baseline results on a Libriphrase hard dataset, increasing AUC from $88.52$ to $94.9$ and reducing EER from $18.82$ to $11.1$.

\end{abstract}


\section{Introduction}
\label{sec:intro}

\begin{figure*}[t!]
  \centering
  \includegraphics[width=\linewidth]{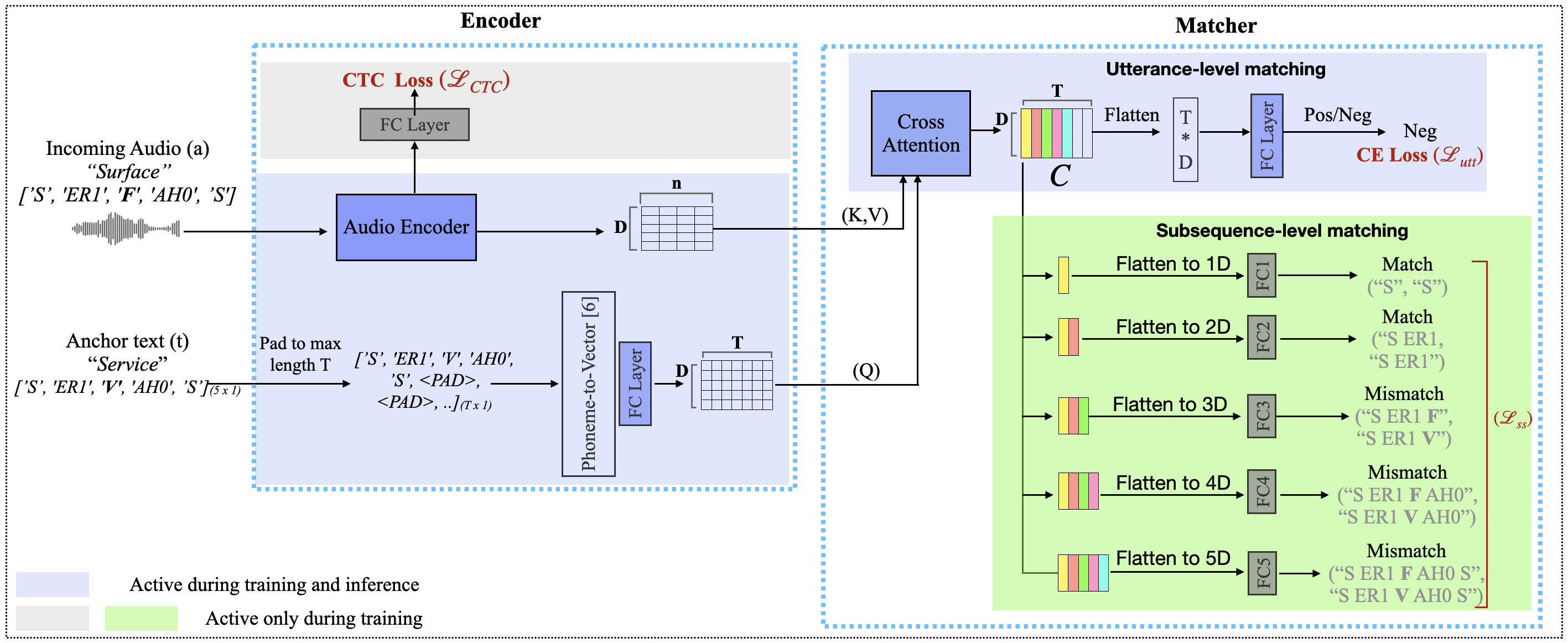}
  \caption{Overall architecture of the proposed method SLiCK (exploiting Subsequences for Length-Constrained Keyword spotting). The model is trained with a multi-task learning approach: \textbf{Matcher} (utterance-level matching $\L_{utt}$, novel subsequence-level matching $\L_{ss}$) and \textbf{Encoder} (phoneme recognition $\L_{CTC}$). We only use the blue parts for inference.}
  \label{fig:slick}
\end{figure*}




KeyWord Spotting (KWS) plays a crucial role in modern voice-activated systems and smart devices. KWS tasks are broadly scoped for either a pre-defined set of keywords \cite{fixed_kws1, Bittar2023ImprovingVK} or a user-defined set of keywords \cite{Nishu2023FlexibleKS, nishu2023matching, flexible_kws1_qbe3, flexible_kws2_asr_embedding}. With the growing emphasis on personalized and user-friendly devices, the demand for user-defined KWS is higher than ever. Deploying such systems on edge devices ensures privacy and reduces power consumption. However, achieving zero-shot user-defined keyword spotting  with a minimal model footprint \cite{Shin2022LearningAA, Lee2023PhonMatchNetPZ} presents a significant challenge.

Previous approaches have registered user-defined keywords as audio recordings, query-by-example (QbE) KWS methods \cite{qbe1,qbe2, qbe4}. However, this enrollment process can lead to a cumbersome user experience, and the accuracy heavily relies on the consistency of the enrolled audio and the test samples. In response to these disparities, recent works have adopted text-based keyword enrollment process \cite{nishu2023matching, Nishu2023FlexibleKS, kwsusingtts, sdc_feature} and explored audio and text embedding based approaches for user-defined KWS task. For example, two independently trained audio and text encoder were explored to semantically align their embedding in a projected latent space \cite{nishu2023matching}. To reduce the discrepancy between audio and text modalities, an audio-compliant text encoder was developed in \cite{Nishu2023FlexibleKS}. In \cite{adaptiveinstance}, the text encoder was biased to produce keyword-conditioned normalization parameters to process with the input audio. Some works have investigated multi-modal KWS \cite{Ai2024MMKWSMP}, combining both audio and text enrollments with integrated embedding of phoneme, semantic text, and the anchor audio. While these methods have shown promising results, they are not suitable for edge devices due to their increased model parameters and overall large package size.

Our focus is on KWS task in low-resource settings, which aligns closely with \cite{Shin2022LearningAA, Lee2023PhonMatchNetPZ}. Some approaches investigated explicit audio-text patterns using monotonic matching loss \cite{Shin2022LearningAA}. But such fixed patterns compromise the alignment flexibility, and may lead to poor performance for similar-sounding keywords. To address this challenge, \cite{Lee2023PhonMatchNetPZ} introduces an auxiliary  phoneme-level detection loss. Nonetheless, detecting phonemes in isolation lacks context and could be sub-optimal for distinguishing similar-sounding keywords \cite{Fang2020UsingPR}.

The complexity of a KWS task gets exacerbated when processing the keywords of varying lengths. To manage this variability, prior works have employed different aggregation methods: using recurrent neural network layers and consuming only the last output \cite{Shin2022LearningAA, Nishu2023FlexibleKS}; applying simple embedding average over the variable sequence length \cite{flexible_kws2_asr_embedding}; and exploring a dynamic sequence partitioning algorithm to effectively align audio-text together \cite{nishu2023matching}. These methods cause information loss during aggregation, reducing model accuracy.



To address the challenges, namely context-lacking phoneme detection and context-losing aggregation, we propose SLiCK, a multi-task learning approach where we effectively manage length variability by imposing a practical constraint on the maximum supported keyword length. In a nutshell, SLiCK comprehensively trains for the KWS task by: 1) conducting an utterance-level matching of the anchor text with the audio, 2) matching the subsequences of the anchor text with the spoken content, and 3) recognizing phonemes in the incoming audio. We make the following contributions:

\begin{itemize}
    \item We propose a novel subsequence-level matching scheme to enhance the model's ability to learn fine-grained audio-text relations. This scheme enables the model to distinguish between similar-sounding keywords through enhanced context.
    
    \item We introduce a length-constrained approach for KWS that preserves context by avoiding explicit aggregation over the variable text length, instead applying a practical constraint on the maximum keyword-length supported.

    \item We design a multi-task training framework that performs comprehensive audio-text matching at various granularity: utterance-level matching, subsequence-level matching, and phoneme recognition.
\end{itemize}

\section{Proposed Method}
\label{sec:proposed_method}

In this section, we describe our proposed method, SLiCK (exploiting Subsequences for Length-Constrained Keyword spotting), shown in Fig. \ref{fig:slick}. The objective is to check if the user-defined keyword is present in the audio. The challenge is that the keyword could have any arbitrary length. Our method takes the incoming audio and the keyword text as input and produces a matching score. SLiCK comprises of two modules: \textbf{Matcher}, and \textbf{Encoder}. The model is trained using a multi-task learning paradigm, simultaneously tackling three tasks: $1$) utterance-level matching task, $2$) novel subsequence-level matching task, $3$) phoneme recognition from the input audio. After training, we retain only the layers related to the utterance-level matching task (blue part in Fig. \ref{fig:slick}) and discard the rest, making the model lightweight for deployment.


\subsection{Encoder}
\label{subsec:Encoder}

\hspace{-0.0in }\textbf{Audio encoding:} 
We process the incoming audio using a tiny conformer \cite{Gulati2020ConformerCT} as the audio encoder, similar to \cite{nishu2023matching, Nishu2023FlexibleKS, Ai2024MMKWSMP}. The audio can vary in length, so the audio encoder produces an embedding of dimension $n \times D$, where $D$ is the embedding dimension and $n$ is the variable time dimension corresponding to the audio length. This embedding vector is used as key $K$ and value $V$ in the subsequent Matcher module. We use a linear layer on top of the audio encoder and perform the phoneme recognition task using the CTC loss \cite{Graves2006ConnectionistTC} $\L_{CTC}$.
\smallskip

\hspace{-0.2in }\textbf{Length-constrained keyword spotting:} 
We address the challenge of supporting user-defined keywords of variable text length by imposing a practical constraint on the maximum supported length. As shown in Fig. \ref{fig:slick}, we accept the keywords of varying lengths and pad them to the maximum supported keyword-length $T$ for further processing, where keyword-length is defined as the number of phonemes in the keyword. 

\begin{figure}[t!]
  \centering
  \includegraphics[width=\linewidth]{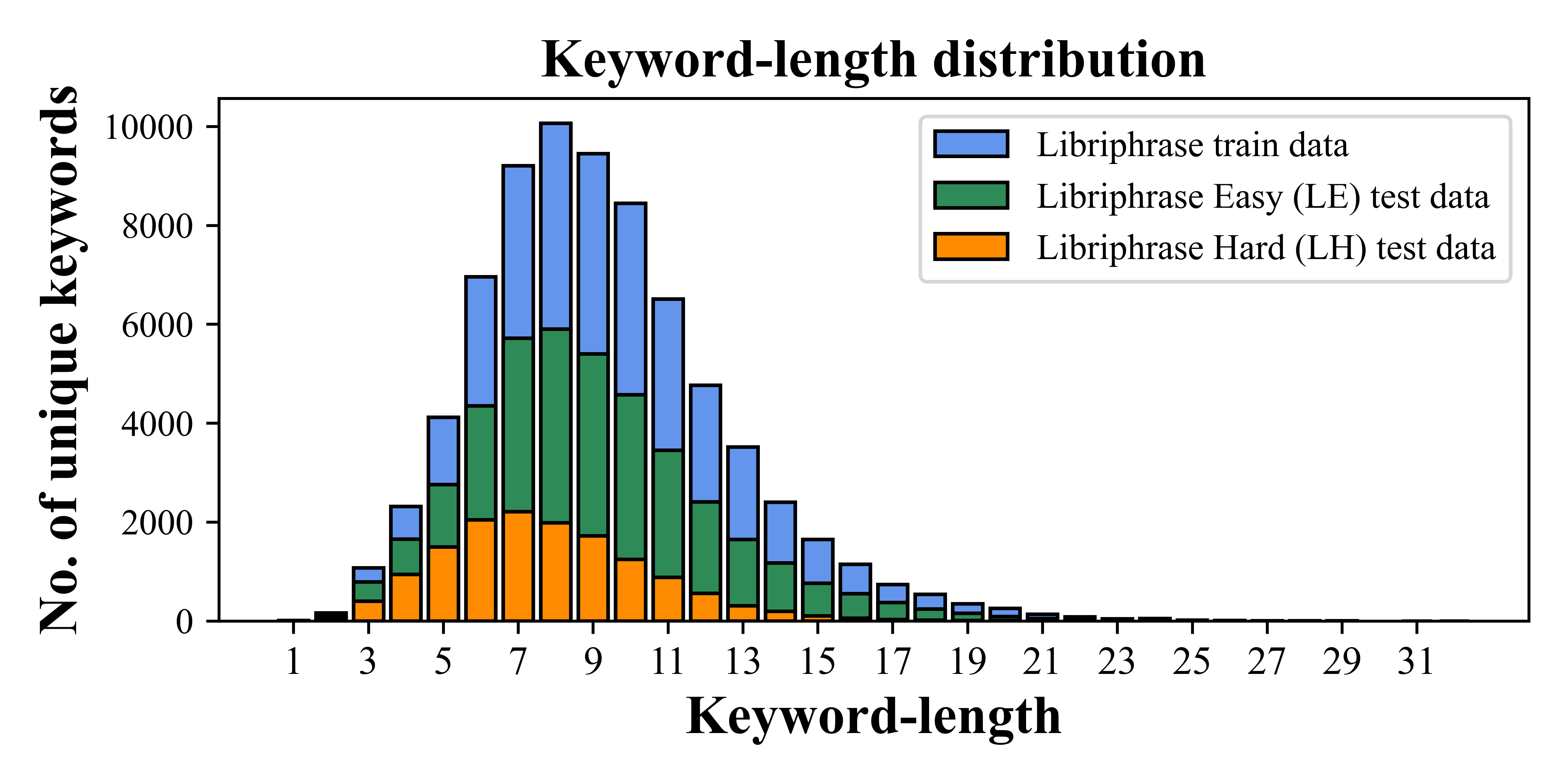}
  \caption{Visualization of keyword-length (no. of phonemes) and their commonality in the datasets. Nearly all keywords are bounded by a maximum length of $25$.}
  \label{fig:keyword-len}
\end{figure}

We analyze the keywords in the Libriphrase dataset \cite{Shin2022LearningAA} and summarize the results in Fig. \ref{fig:keyword-len}. The plot shows the histogram of keywords in terms of the keyword-length (x-axis) and their commonality (y-axis) in all the datasets. We find that there is a wide variety of keywords of varying keyword-length. And nearly all the keywords have keyword-length fewer than $25$ phonemes and majority of them are concentrated around the middle. These insights also generalize well with other popular keyword spotting data \cite{google_data, Qualcomm_data}. We provide sample keywords in the Table \ref{table:keyword samples}. For instance, a keyword with a length of $25$ phonemes, such as ``called the philosophic standard", is considered very long and generally sufficient as the maximum length for a keyword spotting task. These findings led us to approach user-defined keyword spotting as a length-constrained problem with a practical bound on maximum keyword-length supported. This strategy eliminates the need for explicit aggregation over variable text lengths, enabling high accuracy with minimal model footprint. We keep $T=25$ for all our experiments. 
 

\smallskip
\hspace{-0.2in }\textbf{Text encoding:}
The user-defined keywords are enrolled as text, and we convert the text graphemes (spelling) to phonemes (pronunciation) using a pre-trained G2P (grapheme-to-phoneme) system \cite{g2pE2019}. The phoneme sequence is passed to the model as input anchor text which can be of variable length. For instance, the anchor text `\textit{Service: [S, ER1, V, AH0, S]}' is of length $5$. Hence, we first pad the anchor text up to the maximum supported length $T$. We prepare a phoneme-to-vector (P2V) database \cite{Nishu2023FlexibleKS} mapping to get the phoneme embedding vector for each phoneme . We use this mapping to convert the padded anchor text from $T \times 1$ to a text embedding of $T \times D$, matching the embedding dimension $D$ of the audio encoder. This is followed by a linear layer. The text embedding is passed as query $Q$ to the subsequent Matcher module.

\subsection{Matcher}
\label{subsec:Matcher}
 \hspace{-0.0in }\textbf{Utterance-level matching:}
Inspired by the success of transformers, we use a transformer block with $\crossattention$ \cite{Vaswani2017AttentionIA} layers in our Matcher module, calculated as,
\begin{equation}
    C \coloneqq \crossattention(Q,K,V) = \softmax \bp{ \frac{QK^T}{\sqrt{D}} } V,
\end{equation}
with $Q\in\R^{T\times D}$, $K\in\R^{n\times D}$, and $V\in\R^{n\times D}$. The advantage of our novel length-constrained keyword spotting is that the dimension of the attention output $C$ becomes independent of the anchor text length and is always $T \times D$. Hence, unlike \cite{nishu2023matching, Lee2023PhonMatchNetPZ}, we do not require any aggregation over the sequence dimension, before classification. Without any loss of information, we flatten the output $C$ to a vector of size $T\cdot D$, followed by a linear layer to perform the utterance-level matching task. We denote the corresponding cross-entropy loss as $\L_{utt}$.
\smallskip

 \hspace{-0.2in }\textbf{Subseqeunce-level matching:}
Our novel subsequence-level matching scheme enhances the model's ability to learn audio-text relations at a finer granularity and better distinguish similar-sounding keywords through enhanced context.



Let us denote the attention output $C\in\R^{T \times D}$ as a sequence of vectors in $\R^D$ such that $C=\bp{c_1, c_2,\ldots, c_T}$. Let $C_{1:t} \in \R^{t \times D}$ denotes a sub-matrix of $C$ containing the first $t$ vectors, such that $C_{1:t}=\bp{c_1, c_2, \ldots, c_t}$, for any $t\in[T]$. We train the model using $C_{1:t}$ to match the incoming audio and the anchor text up to the length $t$.

The key idea is to match the subsequences of the anchor text with the audio. During training, we have access to the transcription of the incoming audio, referred as spoken text. We use the anchor text and the paired spoken texts to generate the ground truth matching labels for all the subsequences of the anchor text. For example, given anchor text \textit{`Service: [S, ER1, V, AH0, S]'} paired with spoken text \textit{`Surface: [S, ER1, F, AH0, S]'}, we generate a set of five subsequences and the corresponding ground truth matching labels as: \textit{([S], [S], match)}, \textit{([S, ER1], [S, ER1], match)}, \textit{([S, ER1, V], [S, ER1, F], mismatch)}, \textit{([S, ER1, V, AH0], [S, ER1, F, AH0], mismatch)}, and \textit{([S, ER1, V, AH0, S], [S, ER1, F, AH0, S], mismatch)}. 

As shown in Fig.~\ref{fig:slick}, we use a fully connected layer `\textit{FC}$t$` for each subsequence of length $t\in[T]$. The matrix $C_{1:t}$ is flattened into a vector of size $t\cdot D$ and pass through `\textit{FC}$t$` for prediction, which is trained using the corresponding matching label and the cross-entropy loss. If the anchor text is shorter than $T$, the loss is aggregated only up to the valid length. The loss across all `\textit{FC}$t$` is denoted as the subsequence loss $\L_{ss}$.

\begin{table}[t!]
\centering
\begin{tabularx}{\linewidth}{
    |>{\hsize=.2\hsize}X|
    >{\hsize=.8\hsize}X|
  }
    \hline
    \textbf{Keyword-length} & \textbf{Keywords} \\
    \hline
    $1$ & \textit{i}, \textit{a} \\
    \hline
    
    $5$ & \textit{service}
, \textit{surface}, \textit{empire} \\
    \hline
    
    $10$ & \textit{conformation}, \textit{institution}, \textit{experience} \\
    \hline
    
    $15$ & \textit{sixteen hundred}, \textit{their constitution} \\
    \hline
    
    $20$ & \textit{when the united states}, \textit{the premises of mister} \\
    \hline
    
    $25$ & \textit{`called the philosophic standard`}, "\textit{a pleasant breezy apartment}" \\
    \hline
\end{tabularx}
\caption{Keywords of varying lengths (no. of phonemes).}
\label{table:keyword samples}
\end{table}

\subsection{Training criterion}
\label{subsec:Training criterion}
Our training employs a multi-task learning approach with the losses $\L_{utt}$ and $\L_{ss}$ as defined in section \ref{subsec:Matcher}, and $\L_{CTC}$ as defined in section \ref{subsec:Encoder}. The total loss $\L$ is formulated as,
\begin{equation}\label{eq:loss}
    \L = \alpha_1\L_{utt} + \alpha_2\L_{ss} + \alpha_3\L_{CTC},
\end{equation}
where $\alpha_1$, $\alpha_2$, and $\alpha_3$ are weights for the three tasks. We found $\alpha_1=2$, $\alpha_2=1$, and $\alpha_3=5$ to be the best configuration.


\section{Experiments}
\label{sec:experiments}

\subsection{Datasets}
We prepare Libriphrase \cite{Shin2022LearningAA} training and test datasets as in \cite{nishu2023matching, Nishu2023FlexibleKS}, with the test set divided into: LibriPhrase-Hard (LH) for challenging negative pairs and LibriPhrase-Easy (LE) for easier pairs. While additional training datasets usually boosts accuracy \cite{Ai2024MMKWSMP, Nishu2023FlexibleKS}, we focus on evaluating the proposed method without incorporating extra training data and evaluate on LE and LH datasets.


\subsection{Training}
The conformer \cite{Gulati2020ConformerCT} used as the audio encoder consists of \textit{\{$4$ encoder layers, encoder dimension of $D=64$, convolution kernel of size $7$, feed-forward expansion factor of $2$, and $4$ attention heads\}}. In the Matcher module, we employ cross attention layers with \textit{\{hidden size of $64$, filter size of $128$, $4$ hidden layers, and $4$ attention heads\}}.  We train using Adam optimizer \cite{Kingma2014AdamAM} and transformer learning rate schedule \cite{Vaswani2017AttentionIA} with a batch size of $1024$ for $300$ epochs. The inference model (blue part in Fig. \ref{fig:slick}) is optimized for efficiency, containing only $596K$ parameters. We experiment in PyTorch using x86 Linux machines with NVIDIA V100 GPUs.

\subsection{Results and Ablation Study}
We compare our method with the prior works \cite{Shin2022LearningAA, Lee2023PhonMatchNetPZ} designed for low-resource settings. Results show that our proposed method outperforms the baselines on both the LE and LH datasets in terms of Area Under the ROC Curve (AUC) and Equal-Error-Rate (EER) metric, shown in Table \ref{tab:result}. On the LH dataset, our method improves the previous best result of AUC from $88.52$ to $94.9$ and reduces the EER from $18.82$ to $11.1$, all while achieving a model size that is $9\%$ smaller. On the LE dataset, it enhances the baseline AUC from $99.29$ to $99.82$ and reduces the EER from $2.80$ to $1.78$.

\begin{figure}[t!]
      \centering
      \includegraphics[width=0.8\linewidth]{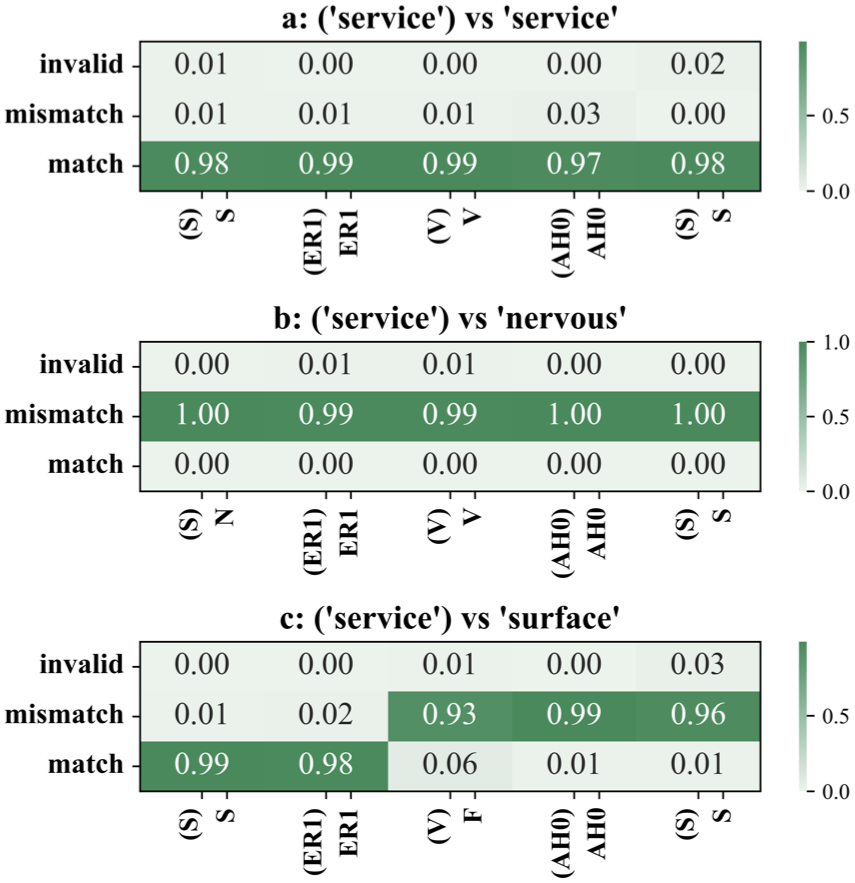}
     \caption{Visualization of predictions from our novel subsequence-level matching scheme for the anchor text \textit{service} across three samples - (a) a positive pair, (b) an easy negative pair with \textit{nervous}, and (c) a hard negative pair with \textit{surface}.}
     \label{fig:subseq scheme}     
    \end{figure}

We visualize the results of our novel subsequence-level matching scheme in Fig. \ref{fig:subseq scheme} for the anchor text \textit{service}. The y-axis represents the ground truth label - match, mismatch, or invalid (for subsequences ending at a padded index). The x-axis shows the anchor text (in parentheses) and the spoken text. Top plot highlights a positive pair, with high confidence for the match labels. The middle plot shows an easy negative pair with strong confidence for the true mismatch labels. The bottom plot shows a hard negative pair, with high scores for the true match labels in the first two subsequences, followed by high mismatch scores for the rest.

We conducted an ablation study to assess the impact of each task in our multi-task training approach. We summarize the results in Table \ref{tab:result}. Training with only the utterance-level prediction task yields an AUC of $88.8$ on the challenging LH dataset. Adding the phoneme recognition task increases the AUC to $92.9$. Finally, incorporating the subsequence-level matching task considerably improves both AUC ($94.9$) and EER ($11.1$) on the LH dataset. We observe a similar trend on the LE dataset.

\begingroup
\renewcommand{\arraystretch}{1.1}
    \begin{table}[!t]          
      \centering
      {
      \setlength{\tabcolsep}{0.35em}
      \begin{tabular}{l|l|l|l|l|l}
        \hline
        \multirow{2}{*}{Method} & \multirow{2}{*}{Param} & \multicolumn{2}{l|}{\textbf{AUC (\%)} $\uparrow$} & \multicolumn{2}{l}{\textbf{EER (\%)} $\downarrow$} \\
        & & LH & LE & LH & LE\\
        \hline
        
        CMCD \cite{Shin2022LearningAA} & 653K & 73.58 & 96.7 & 32.9 & 8.42\\
        PhoneMatchNet \cite{Lee2023PhonMatchNetPZ} & 655K & 88.52 & 99.29 & 18.82 & 2.80\\
         \hline
        \textbf{Proposed} &  \textbf{596K} &  \textbf{94.9} & \textbf{99.82}  &  \textbf{11.1} & \textbf{1.78}\\
         \hline
        
        \hspace{1em} w/o $\mathcal{L}_{ss}$ $^\dagger$ & \multirow{2}{*}{596K} & 92.9 & 99.58  & 14.1 & 2.91\\
       
        \hspace{1em} \hspace{1em}  w/o $\mathcal{L}_{CTC}$ $^\S$ &  & 88.8 & 99.26  & 18.9 & 4.04\\

        \hline
      \end{tabular}
      }
      \vspace{0.1 in}
      \caption{Performance of the baseline methods with similar deployment sizes, proposed method and ablations on Libriphrase Hard (LH), Libriphrase Easy (LE) dataset. $(^\dagger)$ shows the case of training using utterance-level loss and CTC loss. $(^{\S})$ shows the training using just utterance-level loss.}
      \label{tab:result}
    \end{table}
\endgroup

\section{Conclusions}
\label{sec:conclusions}

In this paper, we introduce SLiCK, a length-constrained keyword spotting method based on anchor text subsequences. We tackle the challenge of variable text length by imposing a practical constraint on the maximum length, eliminating the need for aggregation. Additionally, our novel subsequence-level matching scheme enhances the model's ability to learn fine-grained audio-text relations and distinguish similar-sounding keywords. Experimental results show that our method outperforms prior works with a similar model footprint.


\vfill\pagebreak

\bibliographystyle{IEEEbib}
\bibliography{main}

\end{document}